# Femtosecond laser upscaling strategy and biological validation for dental screws with improved osteogenic performance

Mathieu Maalouf [a], Yoan Di Maio [b], David Pallarés-Aldeiturriaga [c], Xxx Sedao [b,c], Lauriane Hivert [a], Steve Papa [a], Elisa Dalix [a], Mireille Thomas [a], Alain Guignandon [a], Virginie Dumas [d].

[a] Université Jean Monnet Saint-Étienne, Mines Saint-Etienne, INSERM, SAINBIOSE U1059, F-42023, Saint-Etienne, FRANCE.
[b] GIE Manutech-USD, 20 rue Benoît Lauras, F-42000 Saint-Etienne, FRANCE.
[c] Hubert-Curien Laboratory, University of Lyon, Jean-Monnet University, UMR 5516 CNRS, F-42000 Saint-Etienne, FRANCE.
[d] Université de Lyon, Ecole Centrale de Lyon, CNRS, ENTPE, LTDS, UMR5513, ENISE, 42023 Saint Etienne, France

**Highlights**

Laser-induced nanostructures with radial orientation shows osteogenic potential.

Large beam reduces texturing time to 40 seconds per dental screw.

180 µm diameter impact renders best mineralization on dental implant surface.

# Abstract

Osseointegration is one of the key conditions for long term successful dental implantation. To this end, titanium alloys undergo plethora of surface treatments able to sustain osteogenic differentiation. For these surface treatments, femtosecond laser (FSL) can generate precise and reproducible surface patterns on titanium, avoiding thermal damage and chemical pollution. We recently identify that laser-induced periodic surface structure (LIPSS) with radial orientation, generated on model (flat) titanium surface, has a high osteogenic potential. However, nano-texturing is time consuming. In the present study, we aimed to reduce the texturing time of radial LIPSS, as well as processing of large commercially available dental screws by ways of laser beam engineering. Our objectives were to maintain at least osteogenic properties demonstrated on model surfaces by adjusting laser beam diameters and to demonstrate maintenance of performance with a dental screw texturing process not exceeding 1 minute.

We first textured model surfaces with radial LIPSS by laser beams of different diameters, with surface impacts of 24µm, 80µm or 180µm, named as R24, R80 or R180 respectively. Osteogenic performance of human mesenchymal stem cells (hMSCs) were compared; seeded on polished control surfaces and textured surfaces and subjected to osteogenic evaluation by cell/matrix imaging, qRT-PCR and mineral deposition quantification. All textured surfaces showed greater osteogenic potential than the control surfaces, with significantly higher efficacy on R180 surface. Therefore, R180 pattern with large beam impacts was chosen for texturing on a dental screw and its osteogenic activity was compared to that of a non-textured screw. Interestingly, R180 required only 40 seconds to be textured on whole screws, on which it preserved a high osteogenic potential. Thus, by using FSL technology, we have improved the osteogenic potential of a topographic pattern while optimizing and scaling up its processing time on a medical device.

# 1. Introduction

In conventional dental implantation surgeries, dental prostheses are most often composed of two parts: a titanium alloy post that is surgically attached to the jaw bone, and an artificial tooth that is supported by its prepositioned base[1]. Despite a decent success rate and a tremendous growth of the dental implant market, up to 19% of dental implantations may face failures[2]. Patients compromised by disease or age are specifically exposed to bacterial infections implying potential loosening, soft tissue deterioration or even bone loss[3,4]. With an aging society, solutions to reduce such failures are crucial to limit health inconvenience, time loss for surgeons or medical expenditures[5].
The current implants market demands high added-value products offering new features at low-cost among which the most desired functionalities are surface texturing of implants to enhance biological functions.

Developing functional surfaces for marketed titanium-based implants for the industry requires actions on multiple levels[6]. Ultrafast laser has been proven, at a laboratory level, as a versatile tool to create various surface nanostructures for rendering a surface with different functionalities [7,8]. For laser microprocessing of implants, the main difficulty comes from the necessity to adapt the laser induced nanometric and micrometric surface features to the macro-scale surface geometries of the implant such as variable windings, lengths and diameters. Consequently, lab-developed processes generally demonstrated on flat surfaces need to be upscaled, both in beam delivery strategies and cycle time to ensure the process industrially viable and cost-efficient. New approaches are necessary in terms of laser parameters and process control, which would logically affect the biological performance of the surface. It is recognized that *in vivo* assessments will provide the necessary biological evaluation of medical products, conforming to ISO 10993 [9]. However, a review of the literature [10] clearly indicates that there is a great reliance on *in vitro* assessments using human mesenchymal stem cells (hMSCs) for both early and late osteogenic gene expressions as well as early Extracellular Matrix (ECM) deposition and ECM late mineralization. Conforming to these *in vitro* standard for implants, we evaluated extracellular matrix production; at 7 and 14 days, we measured the expression of osteogenic genes at 14 days, and finally, at 21 and 28 days, we assessed mineralization which is the final stage in the osteogenic process. This study specifically aims at addressing these challenges by coupling the replication of the biological functions with a more robust and flexible laser process for the industry.

Titanium surfaces textured with linear laser-induced periodic surface structures (LIPSS) through a femtosecond laser (FSL) have shown biological reponses[11,12]. In a previous work, we evaluated whether LIPSS orientation affects osteogenic potential[13]. Interestingly, we found that LIPSS with a radial orientation displayed a greater osteogenic potential than LIPSS with a linear orientation.
We thus propose a strategy for scaling up the radial nanotextures by demonstrating the possibility to achieve complete screw texturing within 40 seconds by increasing the laser beam diameter. The osteogenic performance was evaluated for 3 different diameters of radially oriented nanostructures, with the expectations to get similar results for industrial perspectives. These three textures had impact diameters of 24µm, 80µm and 180µm and were named R24, R80 and R180 respectively. The benefit of deploying a large laser beam is proven to be 3-fold: large impact (namely R180) is found to stimulate most efficiently osteogenesis; a larger beam also facilitates the covering of wider areas within the same amount of time, hence reducing both processing times and unwanted thermal effects. A third advantage lies in the much larger depth of field which enables overcoming the constraints of windings and implant diameter variations, leading to a much more robust process compared to using smaller beams. As a proof of concept, R180 nanotextures was then applied to an entire human dental implant and large-scale cultures using hMSCs for 14, 21 and 28 days were planned to check for osteogenic performance of the textured dental implant as compared to untextured controls. In these upscaled conditions we demonstrated an accelerated mineral deposition at the interface of nanotextured screws as compared to untextured controls, successfully validating efficient texturing processes and biological performance.

## 2. Materials and Methods

### 2.1 Titanium alloy samples

Polished planar samples of TA6V from Goodfellow (Huntingdon, UK) were used in these experiments. The samples had square dimensions of 1 cm² and a thickness of 1 mm, with a roughness of Ra = 0.2 µm.
Ti grade 4 dental screws were sourced from BEGO GmbH (Germany). The head of these prosthetic devices consists of an 11 mm long screw with a core diameter ranging from 4.4 to 2.7 mm.

### 2.2 Laser surface texturing

The surfaces assigned to control group have not been laser-processed and have been referred as UT (Untextured). The micro-machining was carried out using a femtosecond laser (Femto30 from Fibercryst, France). The laser has a central wavelength of 1030 nm with a pulse duration of around 600 fs at full width half maximum and a repetition rate ($R$) tunable from single shot to 1 MHz. The linearly polarized laser pulses were attenuated by a polarizing beamsplitter coupled with a half waveplate, sent through a galvanometric scanner (ExcelliScan14, Scanlab, Germany) and focused through different f-theta lenses ranging from 100 mm up to 340 mm. Prior to the scanner, the laser beam passed through a transmissive S-waveplate (Workshop of Photonics) that aimed at generating either radial or azimuthal polarizations by controlling its orientation regarding the orientation of the input linear polarization (Fig. 1). Positioning of this optical plate was preferentially fixed to get an azimuthal output polarization to favor texturing of radially aligned LIPSS on the workpiece. A second consequence of using such a plate is that the initial gaussian spatial profile is also turned into a donut shape. Adapted beam expander positions were then added on the beam path to get variable beam diameters with donut shapes at the focusing plane.

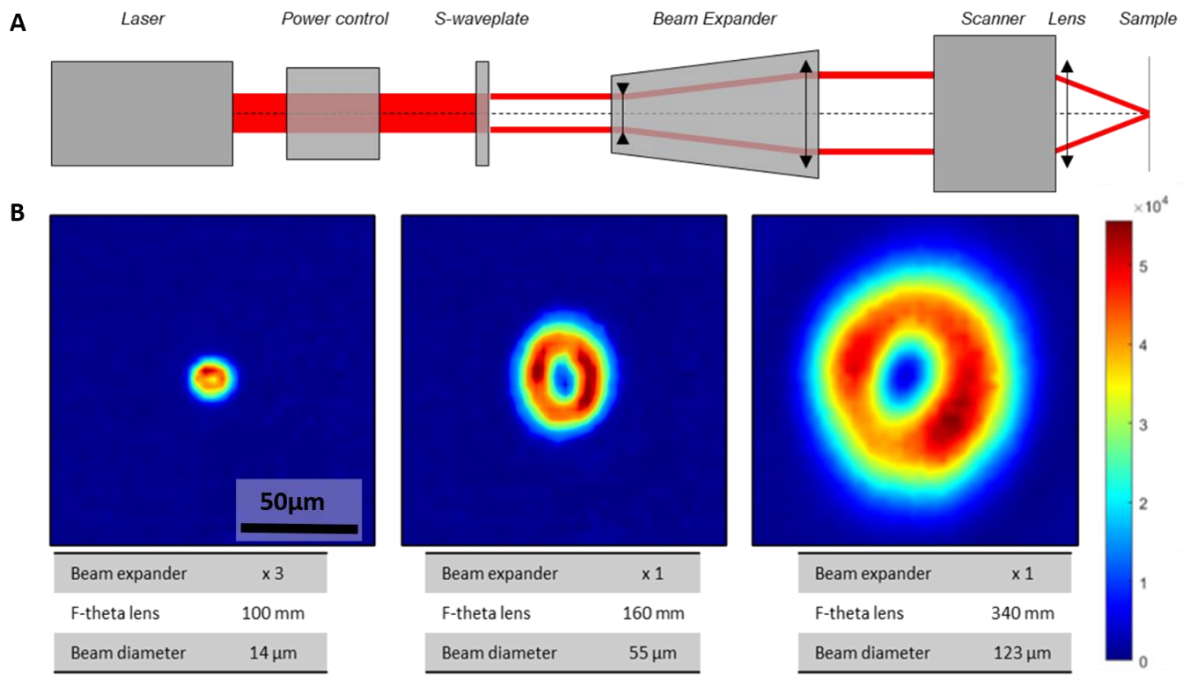

Figure 1: (A) Optical setups. (B) Beam profiles of the vortex beam for diameters 14µm (left) 55µm (center) and 123 µm (right).

Diameters of 14 µm, 55 µm and 123 µm were achieved following the procedure detailed in a previous work[13] and using the configurations given in Table 1. Selected processing parameters respectively led to crater diameters of 24 µm, 80 µm and 180 µm while achieving LIPSS without excessive thermal accumulation. XYZ translation stages enabled the positioning of polished planar samples of TA6V while texturing was managed by high-speed deflection by the scanner. Such planar samples were mainly used for easier and more reproducible biological characterization. Industrial demonstration was finally realized on as-machined dental screws (BEGO GmbH, Germany) that were processed thanks to a complementary rotation system synchronized with laser firing. Texturing was performed by inscribing matrices of N pulses spaced by a value Λ that was determined experimentally searching for the best tradeoff between overlapped surface and LIPSS quality.

Table 1 : Selected laser parameters respectively led to crater diameters of 180 µm, 80 µm and 24 µm.

|  | R24 | R80 | R180 |
| --- | --- | --- | --- |
| Impact diameter | 24 µm | 80 µm | 180 µm |
| Peak fluence | 0.74 J/cm$^2$ | 0.86 J/cm$^2$ | 0.86 J/cm$^2$ |
| Accumulated spots | 10 | 15 | 15 |
| Spacing | 16 µm | 60 µm | 130 µm |

## 2.3 Surface topography analysis

Surface parameters were measured on images taken with an optical 3-D microcoordinate system by InfiniteFocusIF G4, ALICONA (Graz, Austria)[14], which uses the focus variation microscopy technique. The measurements were made with a 1000× magnification, 400-nm lateral resolution and 10-nm vertical resolution. Data topographies were analyzed with Mountains Map® 8.2 software to measure surface roughness parameters focusing on the following: Sa (Arithmetical Mean Height), Sdr (Developed Interfacial Area Ratio) and furrows density and depth. Definitions of surface parameters are stated by ISO 25178 standards.

## 2.4 Cell culture

All titanium samples were autoclaved at 134 °C for 19 min, which is the sterilization procedure of industrial standard to the laser-irradiated samples. hMSCs from PromoCell (hMSC-BM-c, C-12974) at passage 4 were maintained in a T75-flask for 3 days in a growth medium (MSCGM, C-28009, PromoCell). Cells were seeded on titanium flat samples at 7000 cells/cm$^2$ in 24-well plates or on titanium screws at 500k cell/ml in 12-well plates. Both seedings were done with a growth medium. At 24 h post seeding, the growth medium was replaced with an osteogenic medium (MSCODM, C-28013, PromoCell). Thereafter, the osteogenic medium was renewed every 4 days.

## 2.5 Fluorescent cell labeling

Seven days post-seeding, hMSCs cultured in osteogenic medium on flat samples or screws of untextured (UT) and Radial LIPSS textured surfaces were fixed in 10% formalin for 30 min at room temperature. They were then permeabilized with 0.1% Triton X-100 (T8787, Sigma) in phosphate-buffered saline (PBS) for 3 min at room temperature. Samples were incubated with rhodamine-conjugated phalloidin (R415, Invitrogen) diluted at 1:300 in PBS at 37°C for 1.5 h for actin labeling (cytoskeleton). Cells were then incubated with fibronectin (extracellular glycoprotein for cell adhesion) antibody (F3648, Sigma), diluted at 1:100 in PBS, at 37°C for 2 h. Then the samples were incubated with 488 fluoprobes (FP-SA5000, Interchim) diluted at 1:250 in PBS for 1h30 at room temperature. Finally, nuclei labeling was performed with 1µg/ml 4',6-diamidino-2-phenylindole (DAPI, D9542, Sigma) diluted at 1:200 in PBS at room temperature for 20 min. Washes were performed using PBS between each step of the experiment.

## 2.6 Quantitative real time PCR

For qRT-PCR, hMSCs were harvested on UT and Radial LIPSS textured titanium flat samples with Tri-Reagent (Sigma-Aldrich) 14 days post-seeding. RNA amounts were assessed with the Ribogreen kit (Invitrogen, Life Technologies, Eugene, OR, USA) and their quality checked with the TapeStation system (Agilent, Santa Clara, CA, USA). Messenger RNA was reverse transcribed (iScript cDNA synthesis Kit, Biorad) according to manufacturer's instructions, then 300 ng of cDNA were amplified through qRT-PCR using the SYBR Green I dye (Lightcycler faststart DNA masterSYBR green I, Roche, Germany). Primer sequences for the osteogenic genes of interest are given in Table 2. The expression of the housekeeping gene (GAPDH) did not vary significantly within or between groups in either experimental setting (data not shown).

Table 2: PCR primer sequences of genes implicated in different osteogenic pathways.

| Protein | Forward | Reverse | Gene Bank ID |
| --- | --- | --- | --- |
| *ALP* | tgtaaggacatcgcctacca | gaagctcttccaggtgtcaa | NM_000478.5 |
| *COL1A1* | tccggctcctgctcctctta | gttgtcgcagacgcagatcc | NM_000088 |
| *FN1* | ggctggatgatggtagattg | tgcctctcacacttccactc | NM_212482.4 |
| *GAPDH* | catcaccatcttccaggagcga | gtggtcatgagtccttccacga | NM_001289745.1 |
| *OCN* | agcggtgcagagtccagcaa | agccgatgtggtcagccaac | NM_199173.5 |
| *OPN* | tgatggccgaggtgatagtg | atcagaaggcgcgttcaggt | NM_001251830.1 |
| *OSX* | ctggctgcggcaaggtgtat | ccagctcatccgaacgagtg | NM_001300837.1 |
| *RUNX2* | ccttgaccataaccgtcttc | aaggacttggtgcagagttc | NM_001024630.3 |

Abbreviations: ALP: alkaline phosphatase; COL1A1: collagen type 1 α1 chain; FN1: fibronectin 1; GAPDH: glyceraldehyde-3-phosphate dehydrogenase (housekeeping gene); OCN: osteocalcin; OPN: osteopontin; OSX: osterix; RUNX2: Runt related transcription factor 2.

## 2.7 Mineral staining

Mineral staining was performed at 21 days post-seeding on flat samples and at 14, 21 and 28 days post-seeding on dental screws. Cells were fixed in 10% formalin for 30 min at room temperature and then washed with demineralized water. A $3 \times 10^{-3}$ M solution of calcein blue (M1255, Sigma) was deposited on the cells and left for 5 min at room temperature and protected from light. The samples were then rinsed with demineralized water and dried. Of note, calcein blue staining was used to quantify mineralization thanks to its advantage of giving fluorescent images with very low background [15].

## 2.8 Image acquisition and analysis

Images were acquired for all surfaces using a confocal laser microscope (Zeiss LSM 800 Airyscan, Oberkochen, Germany) equipped with Zen software. DAPI and calcein blue labelings were acquired using a 405 nm wavelength laser. Fibronectin labeling was imaged with a 488 nm laser. Actin labeling was acquired with a 561 nm laser.

On flat 2D surfaces, 10 fields of 0.7mm$^2$ were acquired for each condition. These images were analyzed with imageJ software to obtain cell density, actin area and fibronectin area at 7 days, and relative mineralized area as well as the number of mineralized spots at 21 days.

On the 3D surfaces of the screws, volumes ranging from 0.012 to 0.013 mm$^2$ were acquired within screw threads to observe the markers listed above. Mineralization was assessed on 8 volumes per group by measuring the percentage of mineralized surface and the average height of mineralization spots. These two parameters were multiplied and referred to as the mineral formation rate.

## 2.9 Statistics

Raw and derived data were compared between each surface by performing a Wilcoxon–Mann–Whitney U-test with a significance set at p ≤ 0.05. The comparison between Untextured (UT) surfaces and the three types of Radial LIPSS textured surfaces aimed at evaluating the effect of each texture compared to a standard reference. The comparison between all the radial LIPSS surfaces was performed to assess the effect of the different textured spot sizes.

## 3. Results

## 3.1 Surface characterizations

SEM images reveal clear radial LIPSS in all cases with periodic structures radiating outwards from a central point. Surface parameter analysis showed a similar roughness (Sa) and a decrease in Sdr with the beam impact diameter (Fig. 2). The ripples are denser and less deep as the diameter of the beam increases.

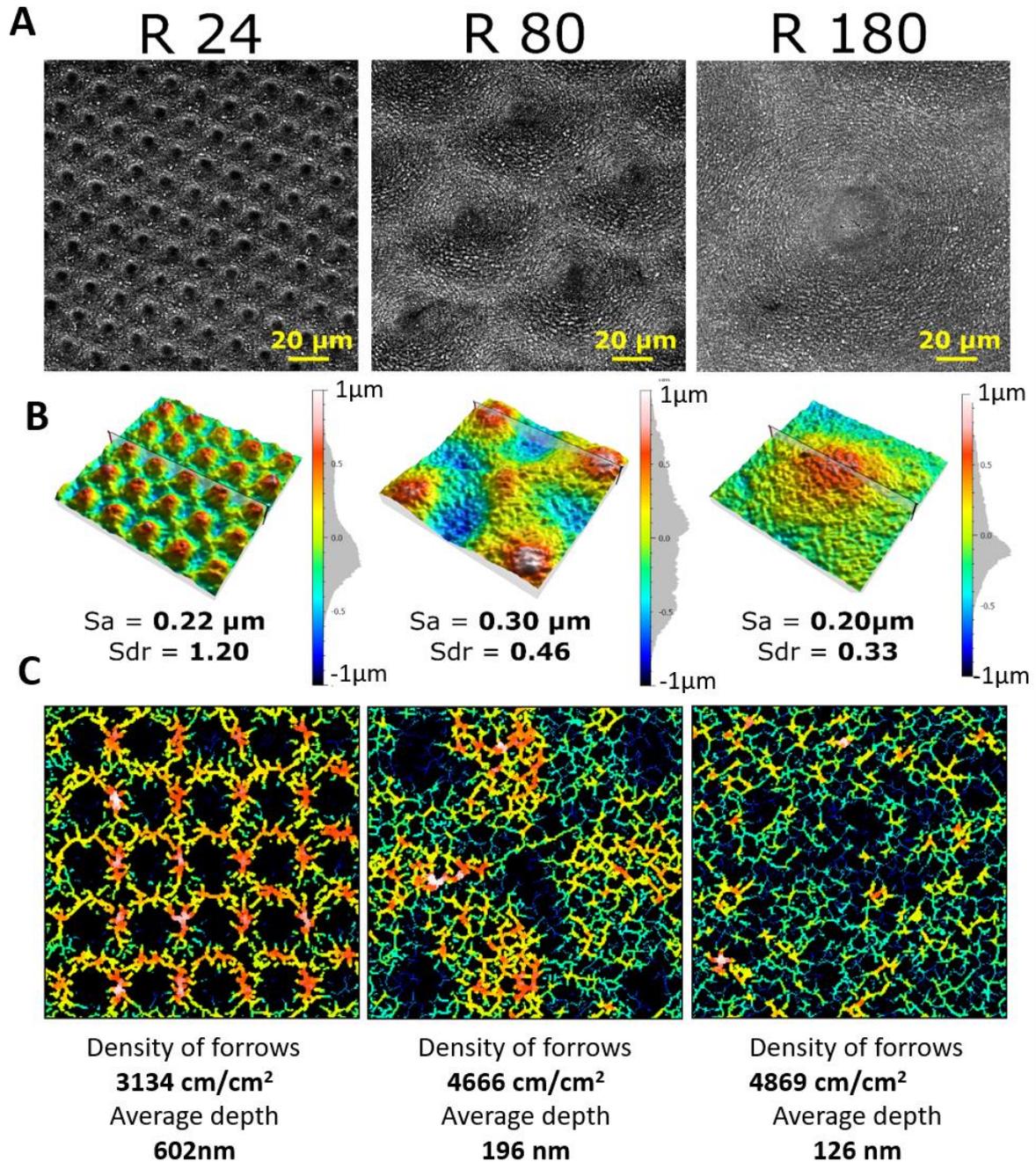

Figure 2 : (A) SEM images. (B) 3D images reconstruction (80µm x 80µm) of the three textures (R24, R80, R180) with radial LIPSS and surface parameter measurements. (C) Representation of furrows density and average depth.

## 3.2 R 180 texturing induces the highest fibrillogenesis on flat samples

To assess the effect of the different Radial LIPSS surfaces on hMSC's extracellular matrix production, we evaluated the fibrillogenesis process (production of fibronectin glycoprotein) at 7 days post-seeding on flat samples. Cell density, calculated by the number of nuclei per titanium surface, was not affected by texturing (Fig. 3). Actin area (spreading of the tissue) was also similar between the 4 surfaces studied (Fig. 3). In contrast, fibronectin area was higher on all textured surfaces compared to UT surfaces (+30% vs R 24, +32% vs R 80, +50% vs R 180). Interestingly, the R 180 surface showed higher fibronectin production than R 24 (+15%, p=0.003) and R 80 (+13%, p=0.0009) surfaces (Fig. 3). No difference was found between R24 and R80 surfaces.

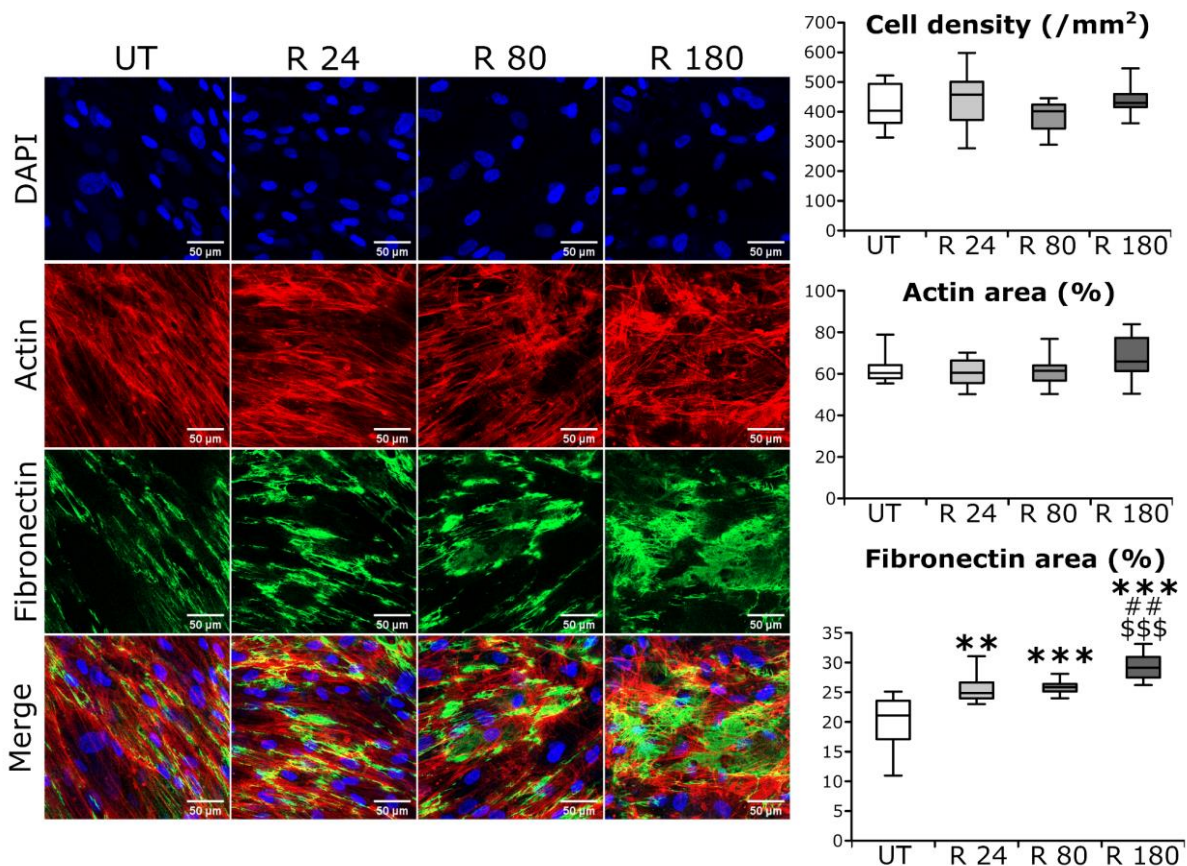

Figure 3: Effect of different Radial LIPSS surfaces on fibrillogenesis 7 days after cell seeding on flat samples. Colored images show DAPI nuclear staining (blue), rhodamine-conjugated phalloidin labeled actin (red), fibronectin antibody coupled with a 488 fluoroprobe (green) and overlaid fluorescent image of immunostained cellular component (merge) for the cells cultured on UT (Untextured), R 24 radial LIPSS, R 80 radial LIPSS and R 180 radial LIPSS surfaces. The graphs show the results of image analysis. **p<0.01, ***p<0.001 vs UT, ##p<0.01 vs R 24, $$$p<0.001 vs R 80, Mann-Whitney U test, n=10 fields/group.

## 3.3 R 180 texturing induces the highest osteogenic-related genes expression on flat samples

To investigate osteogenic-related genes regulation, we performed RT-qPCR at 14 days post-seeding on cells cultured on flat samples. Compared to UT surfaces, osteoblastic differentiation genes RUNX2 and OSX increased in trend on R 24 surfaces and significantly on R 80 and R 180 surfaces (Fig. 4). Matrix production gene FN1 increased significantly on all three textured surfaces compared to UT surfaces. Of note, R 180 surfaces showed a higher FN1 expression compared to R 24 surfaces ($p=0.027$, Fig. 4). Col1A1 gene showed a trend of increase on all three textured surfaces compared to UT surface. For the mineralization genes, ALP was significantly increased on R 24 ($p=0.021$) and R 180 ($p=0.024$) surfaces and in trend on R 80 ($p=0.083$) compared to UT surfaces (Fig. 4). OPN expression increased significantly on all three textured surfaces. Finally, OCN increased only on R 180 surfaces ($p=0.001$) compared to UT surfaces. Interestingly, OCN is significantly higher on R 180 surfaces than on R 24 ($p=0.04$) and R 80 ($p=0.015$) surfaces.

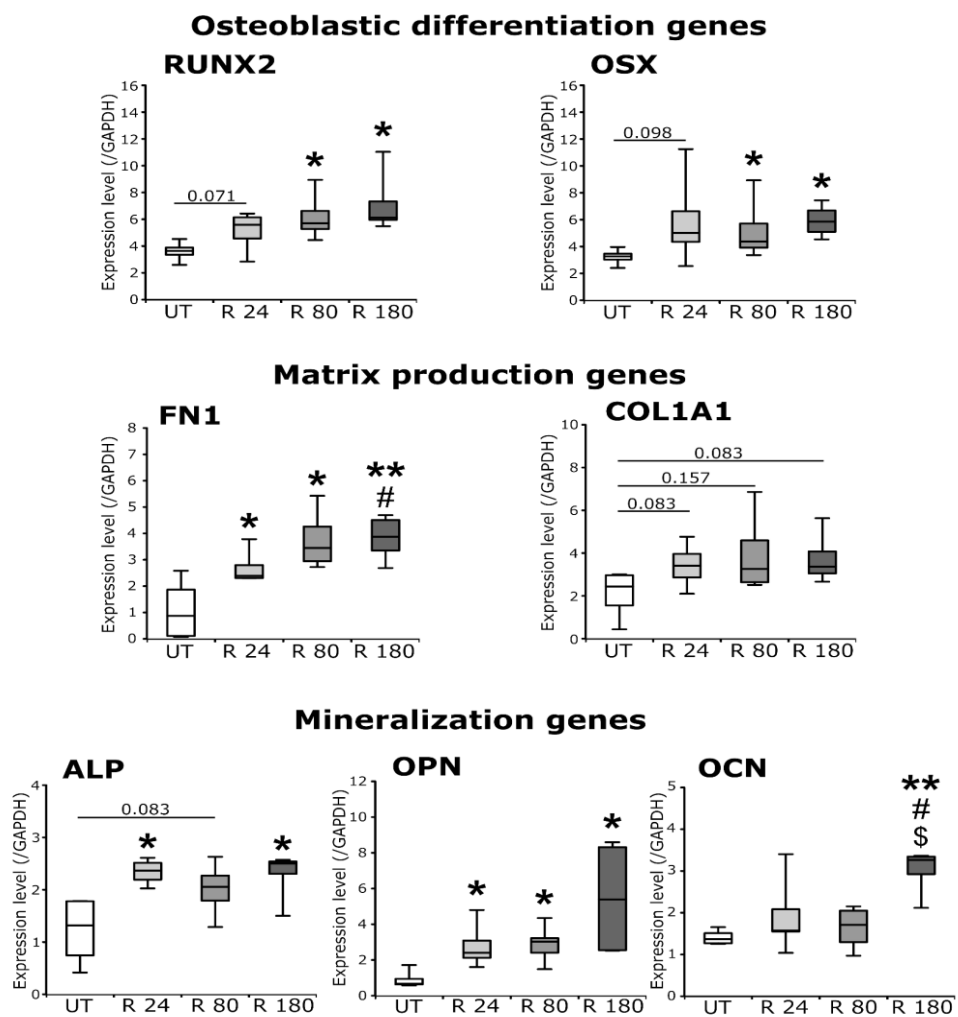

Figure 4: Effect of different Radial LIPSS surfaces on osteogenic gene expression 14 days after cell seeding on flat samples. The graphs show the results of RT-qPCR on the cells cultured on UT (Untextured), R 24 radial LIPSS, R 80 radial LIPSS and R 180 radial LIPSS surfaces. For each graph, values are normalized by GAPDH. *$p<0.05$, **$p<0.01$ vs UT, #$p<0.05$ vs R 24, $p<0.05$ vs R 80, Mann-Whitney U test, n=4-6 samples/group.

## 3.4 R 180 texturing accelerates mineral formation on flat samples

To assess the final step of the osteogenic process, mineralization was assessed at 21 days after cell seeding on flat samples. Compared to UT surfaces, all textured surfaces showed increased mineralization area (+56% for R 24, +147% for R 80, +286% for R 180, Fig. 5A, 5B) as well as higher number of mineralized spots (+83% for R 24, +836% for R 80, +1171% for R 180, Fig. 5A, 5B). Strikingly, the larger the diameter of the Radial LIPSS patterns, the higher the mineralization. Indeed, compared to R 24 and R 80 surfaces, R 180 surfaces displayed larger mineralized surfaces (+147% vs R 24, +56% vs R 80) and higher number of mineralized spots (+593% vs R 24, +36% vs R 80).

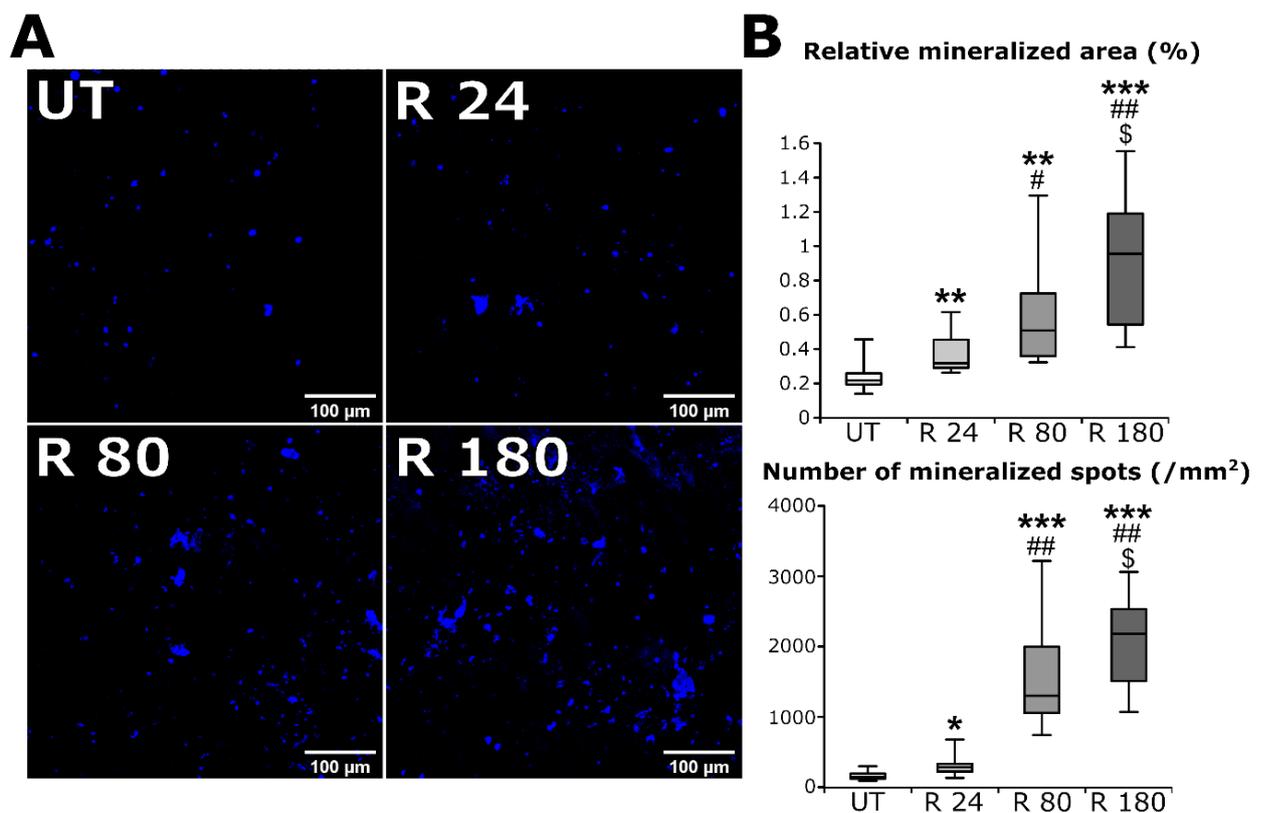

Figure 5: Effect of different Radial LIPSS surfaces on mineral deposition 21 days after cell seeding on flat samples. (A) Colored images show mineralization spots labeled by calcein blue (blue) on UT (Untextured), R 24 radial LIPSS, R 80 radial LIPSS and R 180 radial LIPSS surfaces. (B) The graphs show the results of different quantifications made on the obtained images. *$p<0.05$; **$p<0.01$; ***$p<0.001$ vs UT, #$p<0.05$; ##$p<0.01$; ###$p<0.001$ vs R 24, $$p<0.05$ vs R 80, Mann-Whitney U test, n=10 fields/group.

## 3.5 R 180 texturing on dental screw

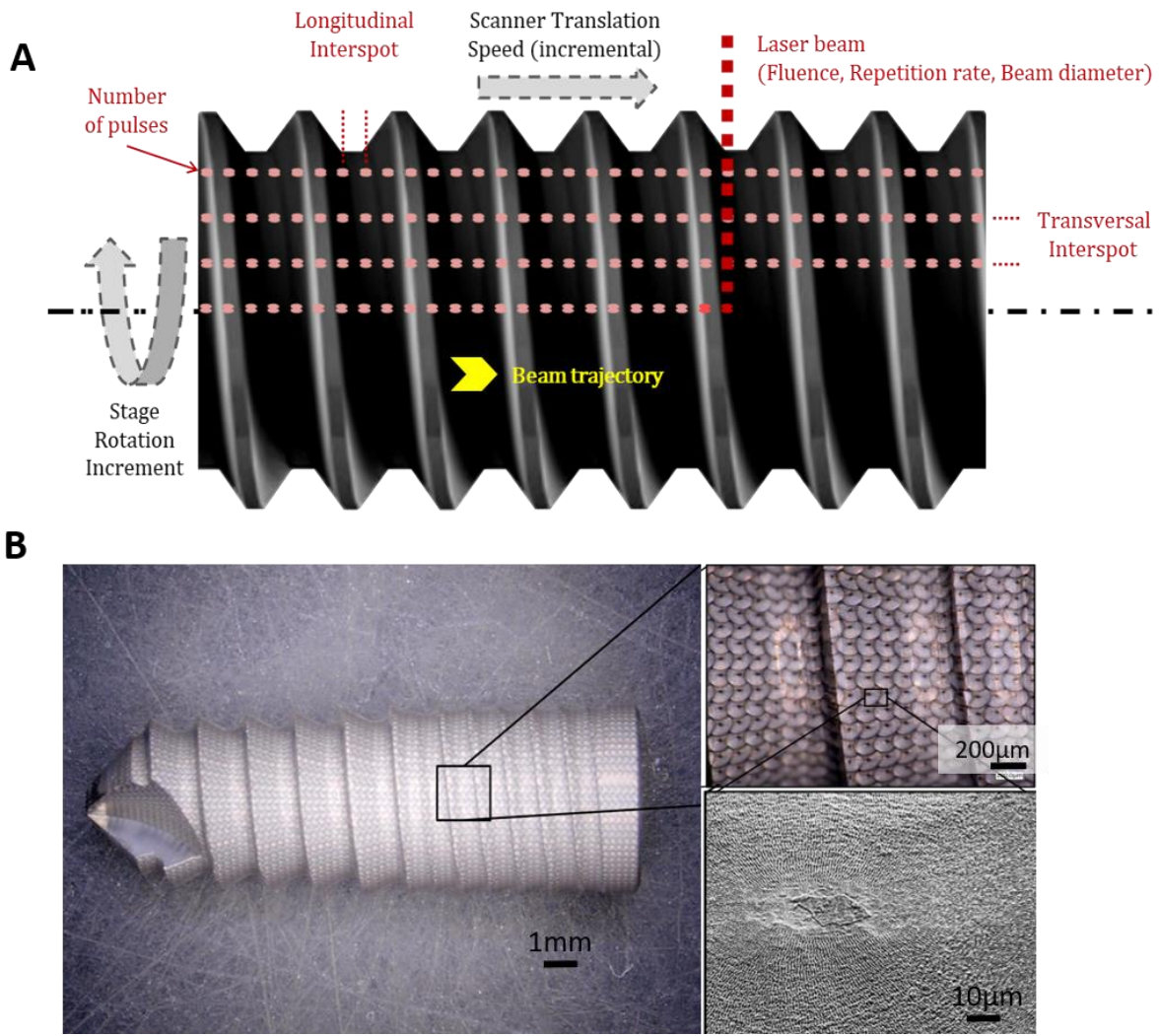

Figure 6: (A) Schematic representation of the laser texturing procedure. (B) Images of dental screw texturing with 180 µm laser impact diameter (radial LIPSS).

In order to test the applicability of the biological results from flat samples to real industrial applications, complete texturing of dental screws was conducted with the best processing parameters, that is to say with 180 µm laser impact diameters, as shown in Fig. 6. Top image describes the processing strategy that was based on the texturing of a complete generator of the screw assimilated to a cylinder before incrementally rotating it. The choice of assimilating the screw as a cylinder was also possible thanks to longest tested focusing lens (340mm) used to generate the 180µm large impacts. Indeed, the estimated depth of field of several millimeters appeared to be much longer than the height of the windings of the screw. This second physical advantage of using larger impacts promoted the robustness of the process independently of a complex mechanical adaptation for the texturing of the screw shape. By doing so, max speed of the scanner was exploited with a repetition rate defined at 50kHz to limit thermal effects while limiting the solicitation of the rotation stage speed.

A total number of 15 pulses was accumulated prior to the translation of the laser beam to the next impact position to maximize the contrast of the LIPSS. Finally, the complete screw was textured within 40 seconds, well below the targeted value of 1 minute, without any significant change of surface quality after laser surfacing when regarding at different positions. Both peaks and valleys between windings are well defined in terms of laser impact contrast and LIPSS are still present as illustrated by the picture at the bottom left and corner, as is shown in the SEM image at the lower right corner of Fig. 6.

## 3.6 R 180 texturing maintains high osteogenic performance on dental screws

To check whether osseointegration is still improved on R 180 Radial LIPSS surface in a medical device model, hMSCs were seeded on UT and R180 Radial LIPSS textured dental screws and their osteoinductive capacities assessed at different time points. At 7 days post seeding, fibrillogenesis was evaluated and showed higher fibronectin production on R 180 texturing (Fig. 7A). Afterwards, mineralization was investigated at 14, 21 and 28 days after cell seeding to check mineral formation dynamic. Mineralized area was more important on R 180 textured screws compared to UT screws, in trend at 14 days post-seeding (+58%, p=0.059) and significantly at 21(+83%, p=0.0008) and 28 days (+42%, p=0.0046) post-seeding (Fig. 7C, D). Mineral thickness, measured as the mean height of the mineral accrual, was higher on R 180 textured screws; in trend at 14 (+27%, p=0.07) and 21 days (+20%, p=0.07) and significantly at 28 days post seeding (+22%, p=0.012, Fig. 7B, D). Mineral formation rate, calculated as the product of mineralized area and mineral thickness, was significantly higher on R 180 textured surfaces at all the time points (Fig. 7D).

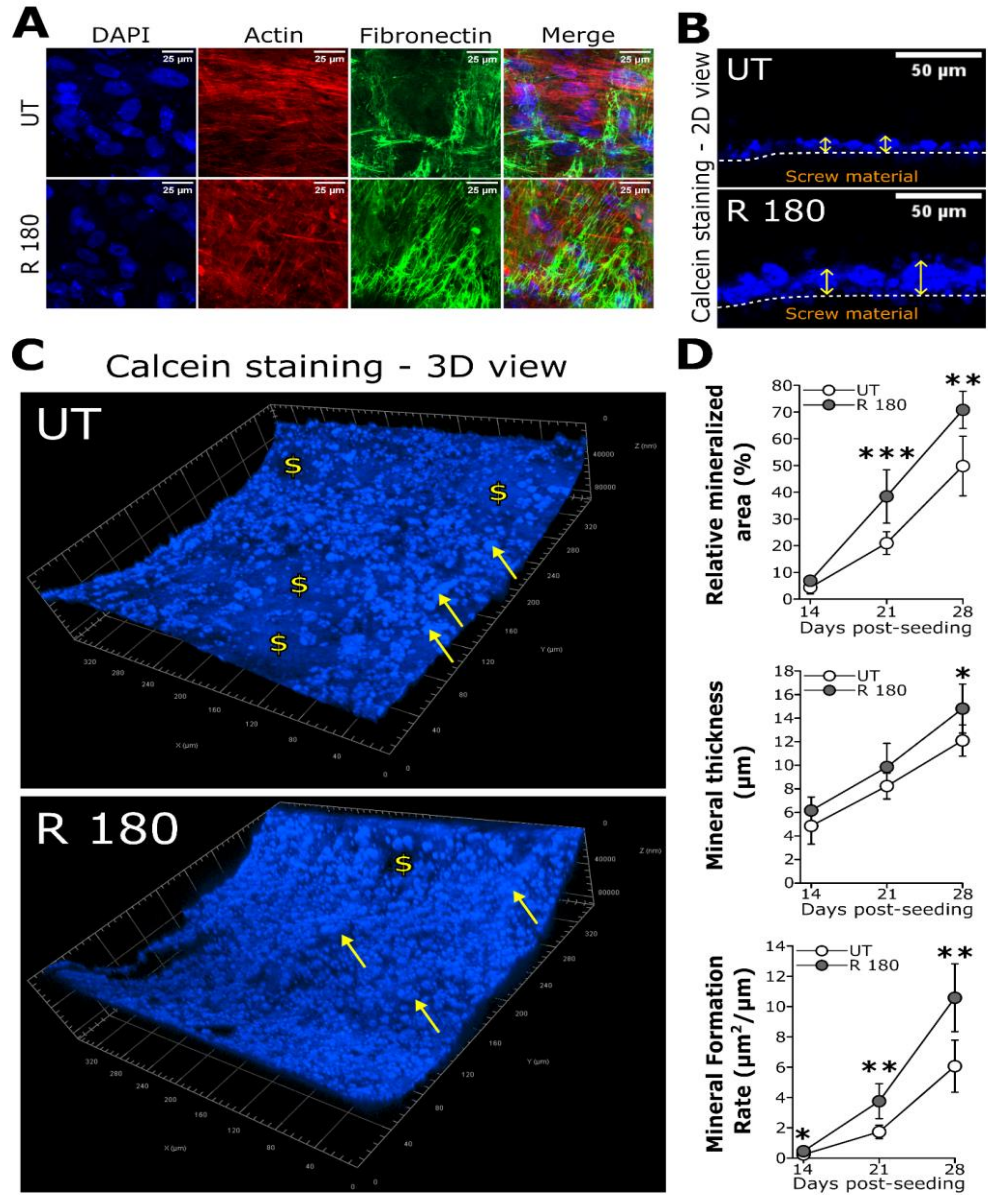

Figure 7: Effect of R 180 texturing on fibrillogenesis and mineral deposition by cells seeded on dental screw. (A) DAPI nuclear staining (blue), rhodamine-conjugated phalloidin labeled actin (red), fibronectin antibody coupled with a 488 fluoroprobe (green) and overlaid fluorescent image of immunostained cellular component (merge) 7 days post-seeding for the cells cultured on UT (Untextured) and R 180 radial LIPSS textured screws. (B) 2D view of mineralization spots labeled by calcein blue (blue) on UT and R 180 radial LIPSS textured screws 28 days post-seeding. Dashed lines delimit screw surfaces. Yellow two-way arrows indicate local mineral thicknesses. (C) 3D view of mineralization spots labeled by calcein blue (blue) on UT and R 180 radial LIPSS textured screws 28 days post-seeding. Yellow arrows point to some mineral spots. $ symbols indicate areas without minerals. (D) The graphs show the results of different quantifications made on the obtained images at different time points. Values are mean ± SD, *p<0.05; **p<0.01 ***p<0.001 vs UT. Mann-Whitney U test, n=8 fields/group.

# 4. Discussion

An ideal dental implant should possess significant potential for achieving osseointegration[16,17]. This mechanism is dependent on efficient mineralization process at the implant surface. However, common biomaterials such as titanium and its alloys, frequently used in clinical dental implants, are bioinert and exhibit limited biological activity[18]. From a biological perspective, early or late implant loss signifies a failure to either achieve or maintain osseointegration. Therefore, enhancing the bioactivity of dental implants is a crucial issue that needs to be addressed[5,19].

Titanium topography plays a key role in enhancing osteogenic cell differentiation on medical implants[20]. The interaction between stem cells and titanium surface patterns have a considerable influence on cell behavior and fate[15,21]. By modulating these surface features, it is possible to direct stem cells towards osteogenic pathways, thereby improving the integration and performance of the implant within the body. The added technical challenge is to ensure reproducibility in the way dental implants are designed, in order to avoid manufacturing hazards and thus minimize biological outcome variations. This is in combination with the need for a non-time-consuming process. To this end, we used the FSL technology to generate osteogenic surfaces reproducibly and rapidly. We followed on from our previous work, in which we identified a highly osteogenic texturing pattern[13]. The strategy we have followed here is to reproduce a similar texturing pattern by increasing the laser beam diameter, thereby reducing the number of laser impacts and thus the texturing time. Once we had identified the best configuration on a 2D substrate, we upscale it to a 3D dental screw surface. Following this approach, we aimed to achieve processing times much less than 1 minute by increasing the laser beam diameter. We focused our work on three different nanostructures obtained by femtosecond laser texturing, which we have labelled R24, R80 and R180. We evaluated the biological response of hMSCs by comparing these different beam diameters, aiming to achieve similar results for industrial applications.

It is well known that FSL-induced topographies can be sensed at the cellular level [22–24] and transduced by the matrix-integrin-cytoskeleton pathway to initiate changes in gene expression[13]. Fibronectin fibrillogenesis, initiating collagen organization, is crucial for osteoblast mineralization. Both the composition and the network organization of the extracellular matrix play an important role in controlling differentiation and mineralization[25]. During mineralization, matrix vesicle are inserted into collagen fibers and their content (bone mineral CaP ie hydroxyapatite crystals) are immobilized between fibers. The organic fibers serve as a scaffolding matrix for mineral deposition and define the size and distribution of apatite crystals[26,27]. In our study, at 7 days, cell proliferation is not affected by surface topography (same cell density) but fibronectin production is stimulated on all textured surfaces with the higher amount found on R180 textured surfaces. Moreover, osteogenic gene expression at 14 days increased on all tested textured surfaces with higher values displayed by cells seeded on R180 surface. The textures therefore make it possible to accelerate osteogenic differentiation. The influence of the different topographies on fibronectin synthesis and assembly affects the overall matrix environment, making it more or less permissive for mineralization. In line with our present data, the early increase of an extracellular matrix of fibronectin at day 7 enhances future mineralization at day 28, which was most effective on R180 surface. Overall, our data suggest that the organization of the fibronectin network is also influenced by the type of texture, the R180 texture certainly allows the creation of a fiber network more favorable to the deposition of apatite crystals compared to the R24 and R80 textures. *In vivo*, bone mineralization involves two distinct phases: primary and secondary mineralization. Primary mineralization in bone occurs through matrix vesicle-mediated processes regulated by osteoblasts. Thereafter, bone mineral density gradually increases during secondary mineralization, presumably due to osteocytic functions[26,28]. At day 28 *in vitro*, the osteocyte differentiation stage is likely not reached, so an *in vivo* study is needed to better analyze the effect of the topography on secondary mineralization.

Regarding surface characterization, the Sa parameter (Arithmetical Mean Height) extends Ra (arithmetical mean height of a line) to a surface, representing the absolute value of the difference in height of each point compared to the arithmetical mean of the surface. This parameter is generally used to evaluate surface roughness and is the most commonly used in the literature to characterize surfaces and correlate them with cellular behaviors such as differentiation[29,30]. The Sa values of the three surfaces, R24, R80, and R180, were evaluated and found to be very similar, with values of 0.22 µm, 0.30 µm, and 0.20 µm, respectively. These values are slightly below the typical range for enhancing osteointegration; the optimal surface roughness should be about 1 µm to promote the osteogenic differentiation of MSCs [31,32]. The Sdr (Developed Interfacial Area Ratio) is defined as the percentage of additional surface area following texturing compared to an original flat surface. We noticed that Sdr decreased with the increase in beam diameter: 1.20 for R24, 0.46 for R80, and 0.33 for R180. A decrease in Sdr concomitant with the increase in surface osteogenic properties has already been reported in another study [33]. Regarding the furrows, their density increased and their depth decreased with the widening of the laser beam. Taken together, our study indicates that a low Sdr values coupled with dense and shallow furrows appears to be the most favorable for osteogenesis. All these data show that the relationship between surface parameters and the acceleration of osteogenesis is complex and multidimensional[34]. It is well known that other physical properties of the surface may favor the effectiveness of osseointegration, such as the wettability, the zeta potential, and the surface energy [35,36]. To complete the characterization of the impact of radial LIPSS on osteoblastic differentiation, it would be also interesting in the future to study the surface chemistry of TA6V samples to evaluate the oxide layer thickness and oxides composition in time due to a well-known surface maturation almost systematically occurring after FSL texturating[37].

Large laser beam diameters such as the 180 µm tested ones also enable a processing depth of field of several mm that helps overcoming the issue of height variations due to windings and evolutive core diameters at different positions of a typical dental screw. Furthermore, larger areas can be textured at once compared to smaller beams, reducing either the processing time or the rotation speed necessary to fit in the processing time targets. Thanks to this approach, the entire dental screw was textured in just 40 seconds, significantly less than the targeted time of 1 minute. This extremely short texturing time is highly compatible with the future industrialization of the FSL texturing process for dental implants. Moreover smooth laser surfacing obtained by the use of low energy LIPSS formation helps maintaining the surface quality without any noticeable change and limited ablated material at any position of the screw.

FSL processing is an advanced method for generating highly reproducible surface patterns, allowing precise and programmable control[38,39] over the creation of patterns, ensuring high uniformity from one sample to another[40,41]. Surface patterns obtained using such lasers also have the advantage of not "polluting" the titanium surface unlike the use of corrosive chemicals during acid etching. It generates micro and nanostructures by creating ultrafast laser pulses that avoid thermal damage and minimize undesirable effects.

Finally, when choosing the design of dental implants, it is crucial to take into account the biological environment of each part of a dental screw once implanted. In the present study, we focused on the implant-bone contact, for which we determined that radial LIPSS obtained with a large laser beam diameter improved osteogenic capacity. In another study, our team also demonstrated that linear LIPSS with a smaller period (390nm) are the most efficient texture for functionalizing the top of the implant. Indeed, linear LIPSS can enhance gingival fibroblast adhesion and limit bacterial colonization, particularly by *Porphyromonas gingivalis,* bacterium deeply responsible for peri-implantitis[42]. Therefore, the FSL could be the ideal tool for texturing an entire implant with various textures tailored to associate dedicated and localized biological functions (gingival adhesion, antibacterial, or osteogenic).

## 5. Conclusions

This *in vitro* study highlights the potential of upscaling strategies for achieving accurate and efficient texturing of dental implants, offering enhanced functional properties and potentially improving their overall performance. By using a larger beam approach, we demonstrate the ability to cover larger surface areas in the same amount of time, reducing processing times and repetition rates that could potentially cause undesirable thermal effects. Clearly, with an appropriate up scaling, the femtosecond laser, generating quickly a wide variety of patterns, is a promising technology for enhancing the properties and performance of titanium in various industrial and biomedical applications.

## 6. Acknowledgments
Steve PAPA reports financial support was provided by French National Research Agency (ANR) - EUR MANUTECH SLEIGHT - ANR-17-EURE-0026. Mathieu Maalouf reports financial support was provided by European Commission LaserImplant project, grant agreement number 951730.